# First JWST/NIRSpec Spectroscopy of O Stars in the Small Magellanic Cloud

Armin Mang Román[1,2], Peter Zeidler[3], Wolf-Rainer Hamann[1], Lidia M. Oskinova[1], M. J. Rickard[1,4],
S. Reyero Serantes[1], H. Todt[1], J. S. Gallagher, III[5], D. Massa[6], D. Pauli[7], V. Ramachandran[8], E. Sabbi[9], and
A. A. C. Sander[8]
[1] Institut für Physik und Astronomie, Universität Potsdam, Karl-Liebknecht-Str. 24/25, 14476 Potsdam, Germany; amang@astro.physik.uni-potsdam.de
[2] Space Telescope Science Institute, 3700 San Martin Drive, Baltimore, MD 21218, USA
[3] AURA for the European Space Agency (ESA), ESA Office, Space Telescope Science Institute, 3700 San Martin Drive, Baltimore, MD 21218, USA
[4] Department of Physics and Astronomy, University College London, Gower Street, London WC1E 66bT, UK
[5] Department of Astronomy, University of Wisconsin–Madison, 475 N Charter Street, Madison, WI 53706, USA
[6] Space Science Institute, 4765 Walnut Street, Suite B, Boulder, CO 80301, USA
[7] Institute of Astronomy, KU Leuven, Celestijnenlaan 200D, 3001 Leuven, Belgium
[8] Zentrum für Astronomie der Universität Heidelberg, Astronomisches Rechen-Institut, Mönchhofstr. 12-14, 69120 Heidelberg, Germany
[9] Gemini Observatory/NSFs NOIRLab, 950 N. Cherry Avenue, Tucson, AZ 85719, USA
Received 2025 March 17; revised 2025 April 4; accepted 2025 April 11; published 2025 May 6

## Abstract

Determining how much mass is removed by stellar winds is crucial to understanding massive star evolution and feedback. However, traditional spectroscopic diagnostics in the UV and optical are not sensitive enough to characterize weak stellar winds of OB stars in low-metallicity environments. A new tool to access weak stellar winds is provided by spectroscopy in the infrared (IR). Stellar atmosphere models indicate that the hydrogen Brα line at $\lambda$ 4.05 $\mu$m is a useful mass-loss rate indicator, particularly at low metallicity. The unprecedented capabilities of the NIRSpec spectrograph on board the James Webb Space Telescope allow us to measure this line in spectra of massive stars in other galaxies. In this work, we present the first NIRSpec spectra of O-type stars in the Small Magellanic Cloud (SMC), which has a metallicity of only 20% solar. Our sample consists of 13 stars with spectral types ranging from O2 to O9.5, including supergiants, giants, and dwarfs. The stars belong to NGC 346, the most massive young cluster in the SMC. We describe the observing strategy and data reduction, highlighting the treatment of the nebular background emission. The spectra cover the 2.8–5.1 $\mu$m wavelength range, and we detect the Brα line in emission in each of our sample stars. Using a combination of spectral and photometric data ranging from the UV to the IR, we improve the measurements of stellar luminosity and reddening. A first qualitative comparison of the observed Brα line with stellar atmosphere models shows its potential as a wind diagnostic for weak-winded stars.

*Unified Astronomy Thesaurus concepts:* O stars (1137); Small Magellanic Cloud (1468); Stellar winds (1636); Young massive clusters (2049); Stellar luminosities (1609); Stellar atmospheres (1584); Infrared spectroscopy (2285); James Webb Space Telescope (2291)

## 1. Introduction

Massive star feedback is among the main drivers of galaxy evolution (P. F. Hopkins et al. 2014) due to their ionizing ultraviolet (UV) radiation, strong stellar winds, and supernova explosions. In particular, stellar winds play a very important role in the evolution of the star and its circumstellar environment, as they can remove a significant fraction of the stellar mass during its lifetime (A. Maeder & G. Meynet 2000; P. Marchant & J. Bodensteiner 2024; D. M. A. Meyer 2024). However, measuring mass-loss rates ($\dot{M}$) remains very challenging, especially in low-metallicity galaxies. Studying massive stars in metal-poor galaxies is of particular interest as these objects are analogs of massive stars that populated galaxies in the young Universe (P. Madau & M. Dickinson 2014; O. G. Telford et al. 2024; M. Lorenzo et al. 2025).

The standard procedure to characterize stellar winds and determine $\dot{M}$ is to use spectroscopy in the optical and the UV. In the optical, stellar winds manifest themselves as emission in some hydrogen and helium lines (i.e., Hα, He II $\lambda$ 4686), while in the UV, resonance lines of metal ions with P-Cygni profiles, such as C IV $\lambda\lambda$ 1548.2, 1550.8 Å, become visible (D. C. Morton 1967; R.-P. Kudritzki & J. Puls 2000). However, in low-metallicity environments ($Z \leqslant 0.2\,Z_{\odot}$), these wind signatures are very weak or absent, in particular for nonsupergiant stars (J. C. Bouret et al. 2003; V. Ramachandran et al. 2019; M. J. Rickard et al. 2022). This problem is enhanced, not only at low metallicities, by inhomogeneities in the density of the wind, a phenomenon called "clumping" (J. Puls et al. 2008). These inhomogeneities are characterized by the clumping factor, $D$, which indicates the density of the clumps compared to the average density of the atmosphere. Mass-loss rates derived when taking clumping into account result in values 3–10 times lower than without including it (T. Repolust et al. 2004; A. W. Fullerton et al. 2006).

Therefore, new diagnostics are needed to estimate $\dot{M}$ in weak winds and to characterize the clumping structure. Accounting for stellar atmosphere models, the infrared (IR) regime offers a promising diagnostic via hydrogen lines originating from higher energy levels, such as Brα at 4.05 $\mu$m. L. H. Auer & D. Mihalas (1969) predicted that the narrow Doppler core of this line will show up in emission even for hydrostatic atmospheres.

F. Najarro et al. (2011) used the SpeX and ISAAC spectrographs, mounted on the Infrared Telescope Facility

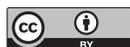

and ESO/Very Large Telescope (VLT), respectively, to obtain spectra of 10 Galactic OB stars in the 0.8–5.4 $\mu$m range. By fitting the Br$\alpha$ line with the non-LTE stellar atmosphere code CMFGEN (D. J. Hillier & D. L. Miller 1998), they constrained $\dot{M}$ for their sample and found agreement with previous results derived from the UV and the optical. They found that Br$\alpha$ reacts inversely to $\dot{M}$. This means that for a fixed set of parameters (temperature, gravity, and luminosity), the emission peak will be higher for stars with a lower $\dot{M}$. Moreover, they found that Br$\alpha$ samples the intermediate region of the wind and can be used to constrain the local clumping factor.

This behavior of the Br$\alpha$ line is attributed to deviations from local thermodynamic equilibrium (non-LTE effects). Low $\dot{M}$ implies a less dense wind, and hence larger departures from LTE (D. Mihalas 1978; A. Lenorzer et al. 2004; N. Przybilla & K. Butler 2004).

However, at the time of their work, instruments were not sensitive enough to obtain IR spectra of stars in other galaxies. Therefore, the potential of Br$\alpha$ as a wind diagnostic for metal-poor massive stars has not yet been tested. With the arrival of the James Webb Space Telescope (JWST), this became possible for the first time. The unprecedented capabilities of this telescope allow us to obtain IR spectra of individual stars in low-metallicity galaxies.

The closest low-metallicity galaxy (with active star formation) is the Small Magellanic Cloud (SMC), a dwarf satellite galaxy of the Milky Way at a distance of 61 kpc (R. W. Hilditch et al. 2005). While close to our own Galaxy, it has a metallicity 5–7 times lower (R. J. Dufour et al. 1982; S. S. Larsen et al. 2000; C. Trundle et al. 2007). The most massive star-forming region in the SMC is the LHA 115–N 66 H II region (K. G. Henize 1956), which is ionized by the young stellar cluster NGC 346 (Figure 1). NGC 346 contains ∼50% of the O-star population of the SMC (P. Massey et al. 1989). Since there is no evidence for supernova explosions in N 66 (C. W. Danforth et al. 2003), NGC 346 provides a clear window into an unpolluted environment with massive stars resembling those in the young Universe.

The OB star population of NGC 346 has been the subject of several spectroscopic studies in the UV and optical wavelength ranges (R. P. Kudritzki et al. 1989; N. R. Walborn et al. 2000; J. C. Bouret et al. 2003; P. L. Dufton et al. 2019). A photometric catalog of objects in N 66 using Hubble Space Telescope (HST)/Advanced Camera for Surveys (ACS) images was presented in E. Sabbi et al. (2007). Recently, M. J. Rickard et al. (2022) analyzed UV HST/STIS and optical VLT/MUSE spectra of 19 O stars in NGC 346 using the non-LTE stellar atmosphere code PoWR (G. Gräfener et al. 2002; W. R. Hamann & G. Gräfener 2003; A. Sander et al. 2015). This work has demonstrated that most metal-poor O-type main-sequence stars have very weak winds such that spectroscopic diagnostics in the UV and optical can provide only upper limits on $\dot{M}$.

Given the exceptional properties of NGC 346, it is not surprising that it was one of the first objects observed with JWST (GTO-1227, PI: M. Meixner), specifically with the Near Infrared Camera (NIRCam) to obtain photometry of the stellar population in the cluster (O. C. Jones et al. 2023; N. Habel et al. 2024). In this Letter, we present the first IR spectra of O stars in NGC 346, taken with the Near Infrared Spectrograph (NIRSpec; P. Jakobsen et al. 2022; T. Böker et al. 2023), all of them being already known and having several parameters determined by previous studies. This makes this sample of stars an ideal target to test the feasibility of JWST to measure weak stellar winds. The NIRCam image of the NGC 346 cluster, with the positions of the targets, is shown in Figure 1.

In Section 2, we describe the observations and the data reduction process. In Section 3, we show the observed Br$\alpha$ line profiles and describe the spectral energy distribution (SED) fitting process that involves multiwavelength photometric and spectroscopic data. Without performing a detailed analysis, we compare the observed Br$\alpha$ line of one target with model profiles computed with different values of $\dot{M}$. Conclusions are drawn in Section 4.

## 2. Observations and Data Reduction

The spectra described in this work were obtained from two different observations (GO-03855, PI: L. Oskinova) with JWST/NIRspec in the fixed slit mode. The instrumental configuration consisted of the S200A1 slit (0.″2 × 3.″2) with the high-resolution G395H/F290LP grating ($R \sim 3000$ at 4 $\mu$m), covering a wavelength range of 2.8–5.1 $\mu$m. Both observations use a 2-point dither with a 3-point spectral nodding pattern to help with cosmic-ray removal and to improve the sampling of the line-spread function.

Observation 1 was obtained on 2024 July 7, using the SUBS200A1 subarray mode with a total exposure time of 150 s. Observation 2 was acquired on 2024 October 6, using the full-frame readout mode with a total exposure time of 1488 s. The full-frame mode allowed us to use the readout pattern NRSIRS2RAPID, which is designed to suppress the detector $1/f$-noise, further increasing the signal-to-noise ratio (S/N; B. J. Rauscher et al. 2017).

The data reduction was done using the 1.16.1 version of JWST Science Calibration Pipeline (H. Bushouse et al. 2025), which includes several packages that process the raw data into fully calibrated science data (CRDS pmap jwst_1298). We used the pipeline steps `calwebb.detector1`, focusing on detector-level correction; `calwebb.spec2`, which performs instrument-specific calibrations on the individual exposures; and `calwebb.spec3`, which is responsible for the combination of the different exposures to obtain the final data product.

The standard pipeline configuration proved to be insufficient to remove large cosmic-ray events (often nicknamed "snowballs") and some detector artifacts. For this reason, we modified the parameters to make the cosmic-ray and outlier rejection algorithms more aggressive, ensuring clean final spectra.

The default pipeline configuration subtracts the background using the spatial dithers. Considering that the targets are surrounded, immersed in an H II region, the background emission can be highly variable in space, making the location of the stars within the cluster relevant for the background-subtraction process. This step is crucial for our objectives since Br$\alpha$ is very prominent in the nebular emission. For the fainter sources and those in a crowded field (see Figure 2), the automatic background subtraction proved to be unreliable. In these cases, we select the background regions directly in the 2D spectrum and perform the subtraction manually.

To check whether the background subtraction was reliable, we measured the velocity of the emission peak in the nebular spectrum. Then, we identify this position in the background-subtracted spectrum and look for signatures of over- or undersubtraction. We show an example of this process in Figure 3. The final reduced spectra are shown in Figure 1.

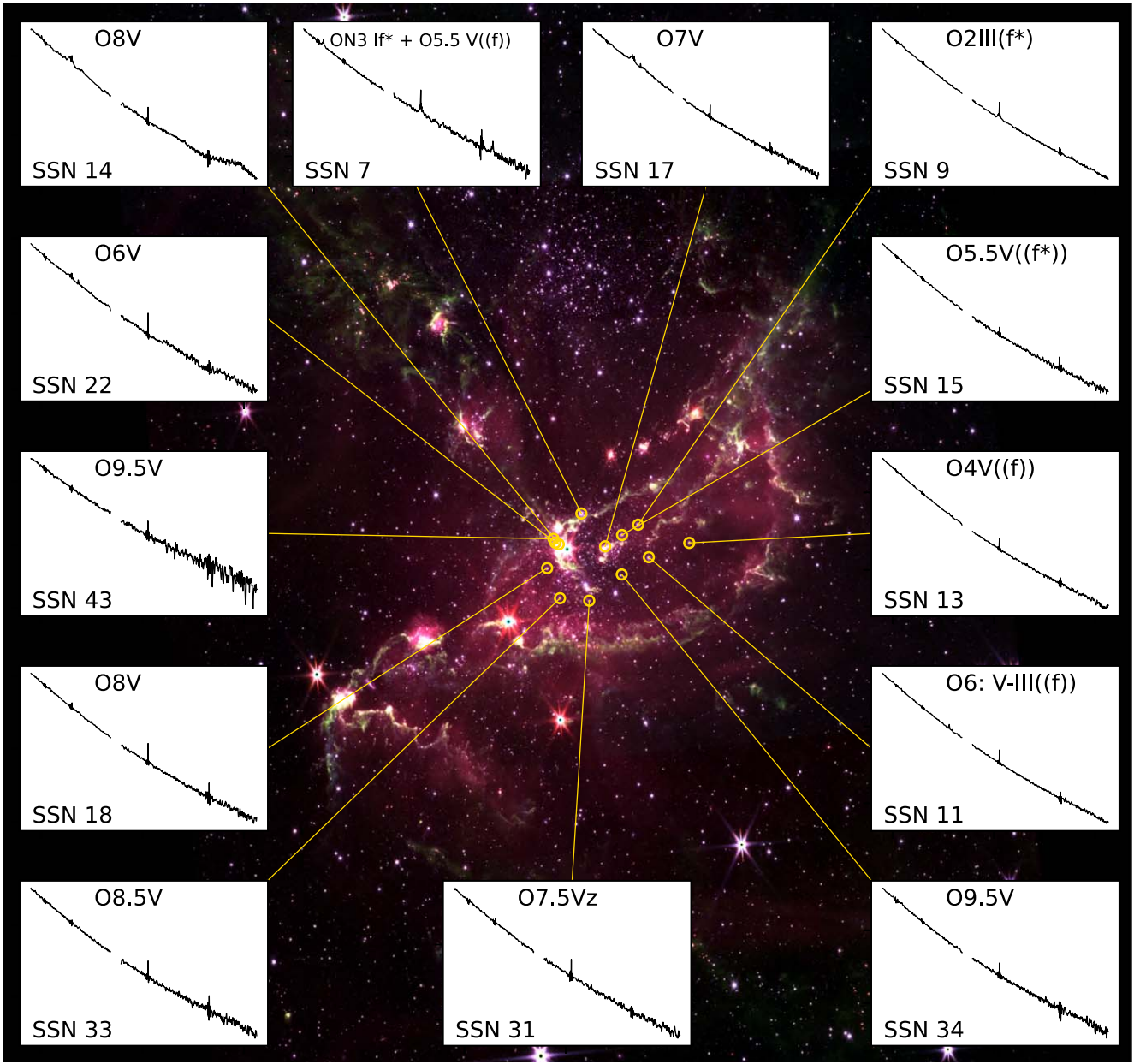

**Figure 1.** NIRCam image of NGC 346 with lines connecting the NIRSpec spectrum to the position of each star of our sample. Catalog names (SSN) come from E. Sabbi et al. (2007), while the spectral types are adopted from M. J. Rickard et al. (2022). The image is a color composite of the F277W (blue), F335M (green), and F400W (red) filters. The broadband filters show the stellar and galaxy continuum emission, while F335M highlights the polycyclic aromatic hydrocarbon (PAH) emission at 3.3 m. North is up, and east is left. The image covers an area on the sky of 5.′4 × 6.′7 that is equal to ∼96 pc × 119 pc at a distance of 61 kpc. The calibrated spectra cover a wavelength range between 2.8 and 5.1 μm range, and the fluxes are between −18 and −16 in log $f_\lambda$ /[erg s$^{-1}$ cm$^{-2}$ s$^{-1}$ Å].

## 3. Results and Discussion

The calibrated spectra around Brα of our sample stars are displayed in Figure 4. The line is in emission in all targets. The line profiles have complex shapes, which might be related to the nebular and background subtraction (see Figure 3). Another factor to take into consideration is binary. Two targets, SSN 7 and SSN 11, are confirmed as multiple systems (M. J. Rickard et al. 2022; M. J. Rickard & D. Pauli 2023).

The peak shows a narrow nebular-like emission and is unresolved (FWHM less than 100 km s$^{-1}$) for most of the fainter targets in the sample (e.g., SSN 31). On the other hand, the peak is broad and resolved for the brightest and earliest stars, such as SSN 7 and SSN 9. There is no obvious trend between the height of the emission peak relative to the continuum and the spectral subtype. The line wings appear to be broad and resolved in all targets, with widths exceeding $\Delta v = 200$ km s$^{-1}$.

All of our targets have available UV and optical spectra and photometry, complementing the new JWST/NIRSpec and NIRCam observations in the IR. This allows for the first time achieving the necessary wide-ranging multiwavelength spectroscopy to calibrate UV, optical, and IR wind diagnostics of low-metallicity O stars.

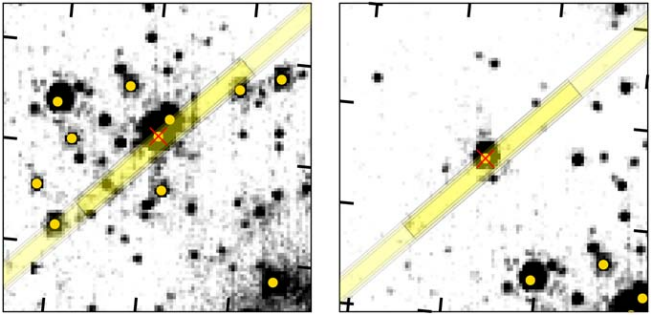

**Figure 2.** Zoom-in (3″.1 × 3″.1) of the F187N NIRCam image for two targets that required manual treatment on the background subtraction. The yellow shaded area indicates the position of the slit in the different dithers, with the darker area being the overlap. North is up, and east is left. SSN 14 (left panel) is located in a very crowded field, with multiple objects falling into the slit (yellow dots indicate JWST/NIRCam catalog sources). SSN 43 (right panel) is the faintest target of the sample and is highly susceptible to spatial background variations.

As an example, we use the O8V star SSN 18. In the top panel of Figure 5, we display the HST/STIS UV spectra used by M. J. Rickard et al. (2022), along with our own new IR JWST/NIRSpec spectra. Additionally, we show the optical HST/ACS photometry from E. Sabbi et al. (2007) and the NIRCam photometry from O. C. Jones et al. (2023). We note that the flux calibration for the NIRSpec spectrum is in perfect agreement with the NIRCam photometry. A synthetic SED, computed with PoWR using parameters from M. J. Rickard et al. (2022), is fitted to the observations. The parameters for the SED fit are the luminosity, $\log L$, and the reddening, $E(B-V)$. M. J. Rickard et al. (2022) derived these parameters based on optical and UV observations only, before the availability of our IR data. By including the IR spectral range, we adjust $\log L$ and $E(B-V)$ to match the observations.

Given the low reddening of our targets, extinction is only marginal in the infrared part of the spectrum. Therefore, we first adjust the luminosity to match the IR observations. The combination of spectra and photometry allows us to check the flux calibration and to see if the slope of the SED is coherent with the observations. With $\log L$ fixed, we then adjust the color excess to match the UV and optical observations. The SED is reddened using the Galactic (M. J. Seaton 1979) and SMC extinction laws (P. Bouchet et al. 1985), assuming a Galactic foreground of $E(B-V) = 0.037 \pm 0.004$ mag (K. D. Gordon et al. 2024). As all other model parameters remain fixed, the uncertainties in our determination of $\log L$ and $E(B-V)$ come only from the fit itself.

After applying this procedure to all targets in our sample, we find that the reddening and luminosity required to fit simultaneously the UV and IR spectra and the optical photometry are systematically lower than the ones obtained without including the IR data (see Table 1 in the Appendix for a comparison with the values from M. J. Rickard et al. 2022). For the case of SSN 18, we find $\log L = 5.08 \pm 0.3$ and $E(B-V) = 0.10 \pm 0.04$ mag. For all targets, we derive a reddening value around 0.08, which is in excellent agreement with the mean reddening of NGC 346 as found by E. Sabbi et al. (2007).

In the bottom panels of Figure 5, we present the observed spectra of the wind-diagnostic lines C IV $\lambda\lambda$ 1548.2, 1550.8 Å (UV), H$\alpha$ (optical), and Br$\alpha$ (IR) for the case of

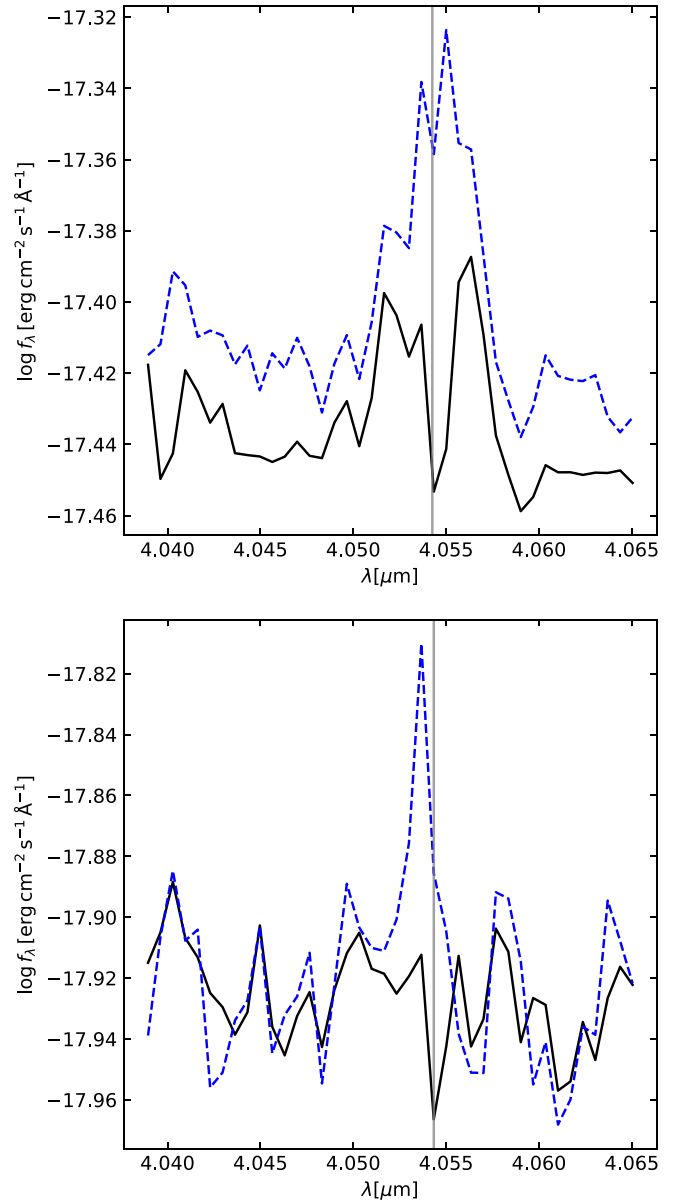

**Figure 3.** Comparison between the background-subtracted spectra around Br$\alpha$ with the default nod-subtraction (solid black line) and manual subtraction (dashed blue line) for two different targets, SSN 14 (top panel) and SSN 43 (bottom panel). The vertical gray line marks the position of the Br$\alpha$ emission peak in the nonsubtracted background. In both cases, the default method oversubtracts the background, while the manual method results in a more reliable line profile.

SSN 18. These observations are compared to synthetic profiles computed with the parameters derived for this star by M. J. Rickard et al. (2022) and different values of $\dot{M}$. On the one hand, the UV and optical wind-diagnostic lines are in absorption as a consequence of the weak wind and, therefore, are not very useful for measuring $\dot{M}$. We see this especially on H$\alpha$, where the models show little reaction to a variation of $\dot{M}$. On the other hand, the Br$\alpha$ line is in emission, agreeing qualitatively with the model prediction. Moreover, we see how the model changes with $\dot{M}$, with the height of the emission peak reacting inversely to it. It is particularly notable that an increase of only 0.05 dex in $\log \dot{M}$ has a very noticeable effect on the height of the Br$\alpha$ line profile. Similarly, we can see that a low mass-loss rate such as $\log \dot{M} = -9.1$ dex shows a high narrow emission peak. A closer inspection shows that $\log \dot{M} \geqslant -8.45$ dex would be required to fit the observed Brα profile. This highlights the sensitivity of the Brα line to $\dot{M}$, especially in weak winds.

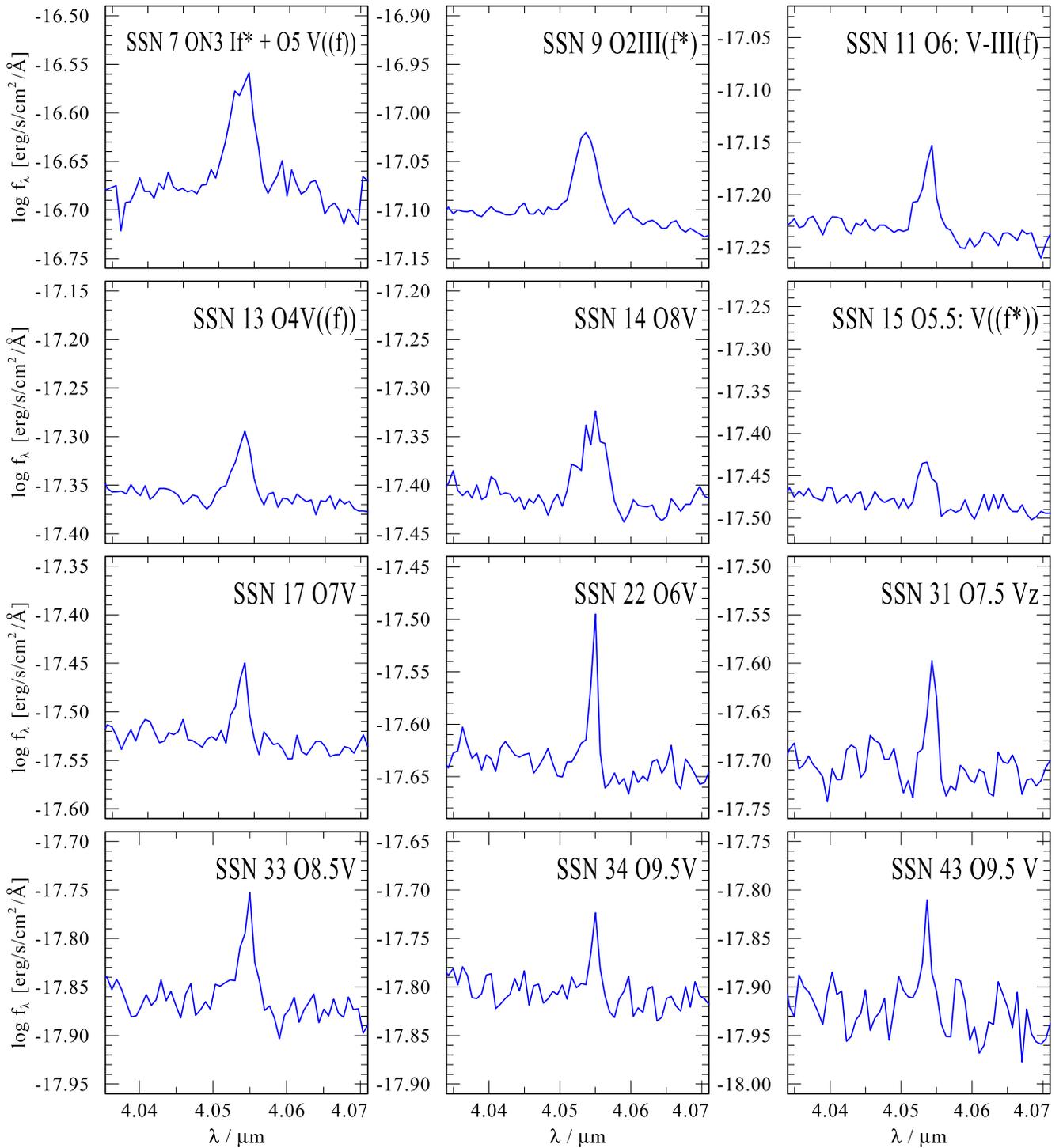

**Figure 4.** Observed Brα line profiles for 12 targets of our sample.

Having underlined the potential of Brα as an IR wind diagnostic, we will start efforts for a consistent spectral analysis of UV, optical, and infrared spectra. Additionally, we will explore in detail the nature of the Brα emission, exploring how the different parameters affect the non-LTE effect causing it. This work, involving many new model calculations with varying mass-loss rates and potentially adjusted stellar parameters, is beyond the scope of the present letter and will be presented in a future publication.

## 4. Summary and Conclusions

Thanks to the capabilities of NIRSpec on board JWST, we obtain, for the first time, NIR spectra of low-metallicity O stars. Our sample consists of 13 stars in the young stellar cluster NGC 346 in the SMC galaxy.

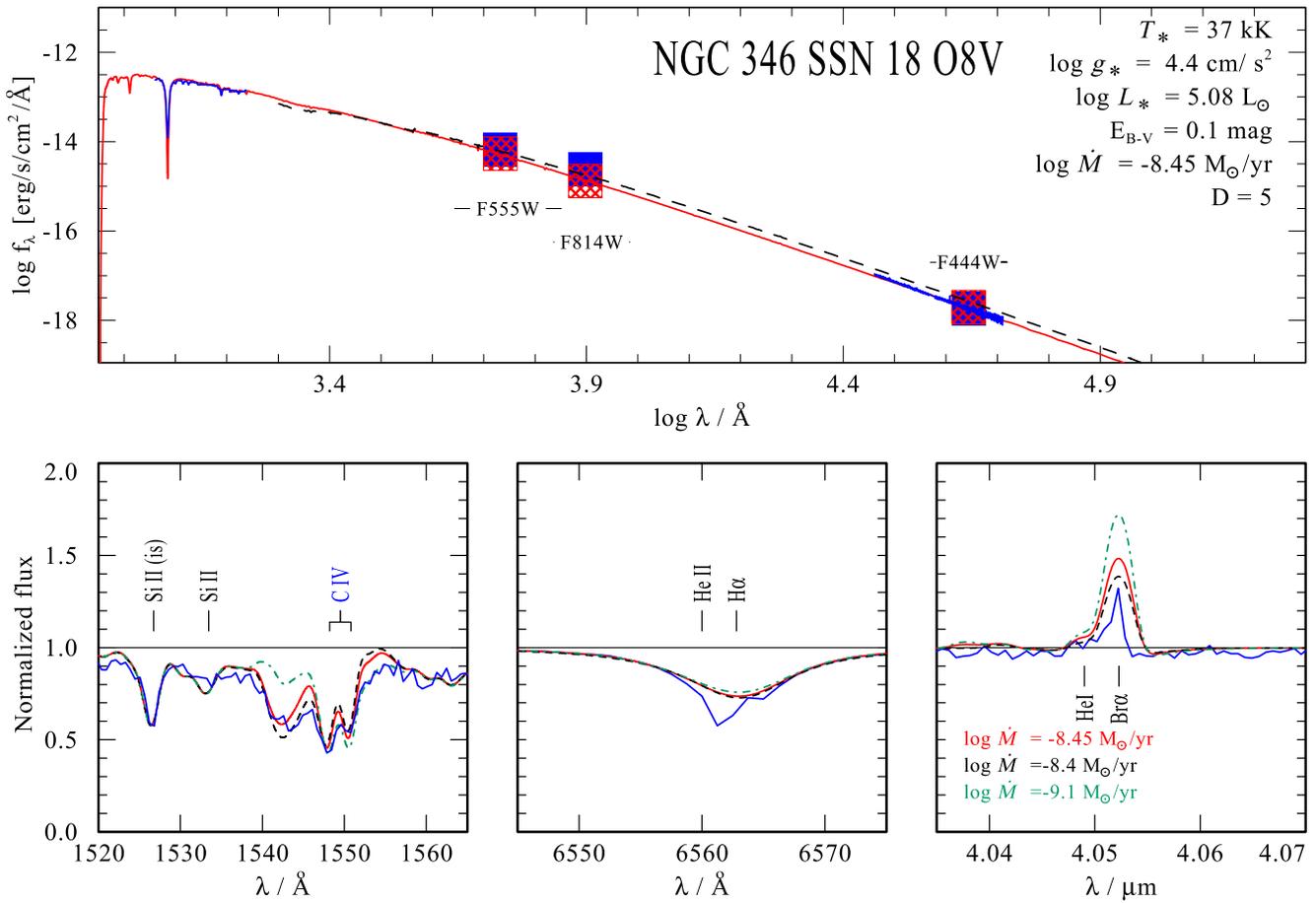

**Figure 5.** Comparison of observations and the stellar atmosphere model computed with parameters derived by M. J. Rickard et al. (2022) for the O8V star SSN 18. Top panel: synthetic SED (red line) adjusted to the HST/STIS UV spectrum, HST/ACS photometry in the F555W (14.470 ± 0.003 mag) and F814W (14.672 ± 0.004 mag) filters, JWST/NIRCam photometry in the F444W (18.786 ± 0.002 mag) filter, and the JWST/NIRSpec spectrum (blue lines and boxes). The red-hatched boxes are centered on synthetic magnitudes derived from the model spectrum. The size of the photometry boxes does not represent uncertainties. The luminosity and reddening indicated in the insert are revised down to match the infrared observations. The black dashed line is the synthetic SED computed with the previous reddening and luminosity values. Bottom panels: observed line profiles of C IV λλ 1548.2, 1550.8 Å, Hα, and Brα, located in the UV, optical, and IR spectra (blue solid lines) compared with models with log $\dot{M} = -8.45$ (red solid line), log $\dot{M} = -8.4$ (black dashed line), and log $\dot{M} = -9.1$ (green dotted–dashed line). The excess absorption in the observed Hα profile is most likely an artifact by an oversubtraction of nebular emission.

We show the importance of not relying on automatic data reduction when analyzing targets located in a crowded region. Specifically, we illustrate how background subtraction is crucial to correctly removing nebular emission.

The main spectral feature is the Brα line, which was proposed as a useful mass-loss rate diagnostic. We detect this line in emission even in the fainter stars in our sample. While we cannot see a clear dependence of the morphology of the Brα with the stellar parameters, a first comparison of an example spectrum to non-LTE stellar atmosphere models shows the potential of this line as a powerful tool to characterize stellar winds. In future work, we will use detailed stellar atmosphere modeling and multiwavelength spectroscopy to further test this line as a wind diagnostic for low-metallicity massive stars.

Furthermore, we show that a multiwavelength approach to SED fitting is required. Model SEDs of our targets derived only from UV and optical data do not match IR observations. To simultaneously fit all three wavelength ranges, a revision of both log L and E(B − V) is necessary. We find that the new revised values are systematically lower than the ones derived in previous studies that did not take the IR range into account.

We conclude that the IR spectral range provides valuable information on low-metallicity O-type stars. The unprecedented capabilities of JWST are needed to develop new strategies to address open problems regarding metal-poor massive stars and their winds.

**Acknowledgments**

This work is based on observations made with the NASA/ESA/CSA James Webb Space Telescope. The data were obtained from the Mikulski Archive for Space Telescopes at the Space Telescope Science Institute, which is operated by the Association of Universities for Research in Astronomy, Inc., under NASA contract NAS 5-03127 for JWST. These observations are associated with program #3855. The JWST data presented in this Letter were obtained from MAST at the Space Telescope Science Institute. The specific observations analyzed can be accessed via doi:10.17909/sgkw-w358.

A.M.R. and S.R.S. acknowledge financial support by the Deutsches Zentrum für Luft und Raumfahrt (DLR) grants FKZ 50 OR 2411 and FKZ 50 OR 2108.

V.R. and A.A.C.S. are supported by the German Deutsche Forschungsgemeinschaft, DFG in the form of an Emmy

Noether Research Group—Project-ID 445674056 (SA4064/1-1, PI: Sander).

E.S. is supported by the international Gemini Observatory, a program of NSF NOIRLab, which is managed by the Association of Universities for Research in Astronomy (AURA) under a cooperative agreement with the U.S. National Science Foundation, on behalf of the Gemini partnership of Argentina, Brazil, Canada, Chile, the Republic of Korea, and the United States of America.

J.S.G and A.M.R, acknowledge support for research in program JWST GO-03885 provided by NASA through a grant from the Space Telescope Science Institute under contract NAS 5-03127 for JWST.

The authors thank the reviewer for the valuable comments that helped to improve the quality of this manuscript.

## Appendix
## Results for Individual Stars

Table 1 gives the individual values and uncertainties for $\log L$ and $E(B-V)$ for all of our targets and compares them to the ones derived by M. J. Rickard et al. (2022). All $\log L$ and $E(B-V)$ values from our work are lower than the ones from the literature. All results are preliminary and might be revised after a fully detailed spectroscopic analysis of the NIRSpec data.

**Table 1**
Derived Luminosities and Reddening for Our Targets Compared with the Ones from M. J. Rickard et al. (2022)

| SSN | $\log L$ (This work) ($L_\odot$) | $E(B-V)$ (This work) (mag) | $\log L$ (M. J. Rickard et al. 2022) ($L_\odot$) | $E(B-V)$ (M. J. Rickard et al. 2022) (mag) |
|---|---|---|---|---|
| 9  | 5.91 ± 0.05 | 0.10 ± 0.03 | 6.12 | 0.13 |
| 13 | 5.45 ± 0.04 | 0.08 ± 0.03 | 5.60 | 0.11 |
| 14 | 5.11 ± 0.08 | 0.10 ± 0.04 | 5.37 | 0.16 |
| 15 | 5.22 ± 0.04 | 0.07 ± 0.03 | 5.45 | 0.12 |
| 17 | 5.10 ± 0.05 | 0.06 ± 0.03 | 5.30 | 0.11 |
| 18 | 5.08 ± 0.03 | 0.10 ± 0.04 | 5.25 | 0.15 |
| 22 | 5.01 ± 0.09 | 0.06 ± 0.04 | 5.32 | 0.13 |
| 31 | 4.95 ± 0.03 | 0.10 ± 0.03 | 5.09 | 0.12 |
| 33 | 4.69 ± 0.03 | 0.08 ± 0.03 | 4.97 | 0.14 |
| 34 | 4.70 ± 0.04 | 0.07 ± 0.03 | 4.97 | 0.13 |
| 43 | 4.61 ± 0.03 | 0.07 ± 0.03 | 4.82 | 0.12 |

## ORCID iDs

Armin Mang Román https://orcid.org/0009-0006-4537-7858
Peter Zeidler https://orcid.org/0000-0002-6091-7924
Wolf-Rainer Hamann https://orcid.org/0000-0002-4970-6886
Lidia M. Oskinova https://orcid.org/0000-0003-0708-4414
M. J. Rickard https://orcid.org/0000-0002-5255-8116
S. Reyero Serantes https://orcid.org/0000-0002-3155-7237
H. Todt https://orcid.org/0000-0002-9684-3074
J. S. Gallagher, III https://orcid.org/0000-0001-8608-0408
D. Massa https://orcid.org/0000-0002-9139-2964
D. Pauli https://orcid.org/0000-0002-5453-2788
V. Ramachandran https://orcid.org/0000-0001-5205-7808
E. Sabbi https://orcid.org/0000-0003-2954-7643
A. A. C. Sander https://orcid.org/0000-0002-2090-9751